\documentclass[5p,twocolumn]{elsarticle}
\journal{Physics Letters B}

\usepackage{graphics}
\usepackage{epsfig}

\begin{document}
\begin{frontmatter}
\title{New estimation for neutron-neutron scattering length: 
charge symmetry and charge independence breaking revisited}
\author{E.S. Konobeevski}
\ead{konobeev@inr.ru}
\author{S.V. Zuyev}
\ead{zuyev@inr.ru}
 \address{Institute for Nuclear Research, Russian Academy of Sciences, 60-letiya Oktyabrya
prospekt 7a, Moscow, 117312, Russia}
\author{V.I. Kukulin}
\ead{kukulin@nucl-th.sinp.msu.ru}
\author{V.N. Pomerantsev}
\ead{pomeran@nucl-th.sinp.msu.ru}
 \address{Skobeltsyn Institute of
 Nuclear Physics, Moscow State University, Leninskie Gory 1, Moscow,
119991, Russia}

\begin{abstract}
Recent experimental results for neutron-neutron scattering length are reanalyzed
from the point of view of three-nucleon force contribution. We found that the
limiting value of $a_{nn}=- 15.8\pm 0.5$~fm must be free of any implicit
three-body force contribution. 
We have also shown that the difference between the above experimental value of $a_{nn}$ and
 the well established value of neutron-proton scattering length $a_{np}$ 
 can be  explained by differences in the one-pion exchange potentials.
\end{abstract}



\end{frontmatter}

\section{Introduction}
 The correct value of neutron-neutron scattering length $a_{nn}$ has a
fundamental importance for nuclear physics at whole as well as for
numerous particular problems like existence of multineutrons, size of
charge-symmetry and charge-independence breaking effects etc. Unfortunately, up to date there
is no clear knowledge on exact value of $a_{nn}$. Many different
values for $a_{nn}$ (in interval from -16 up to -19 fm) which have
been extracted from many different-type experiments must be considered
almost on equal footing (see, e.g.,  
reviews~\cite{Mach_Slaus,Tornow,Chen}). The main
experimental results for $a_{nn}$ have been found
from two different types of experiments:\\
(i) the final-state $nn$ interaction in three-body breakup $n+d \to nn
+p$;\\
(ii) the final-state $nn$ interaction in $d(\pi^-,\gamma)nn$ in stopped
pion radiation capture on deuteron.

The second-type experiments are considered now as most accurate ones due to
absence of three-body rescattering and three-body force effects in the final
state. These experiments have lead to the value $ a_{nn}=-18.9\pm 0.4$~fm
(corrected for the magnetic-moment interaction of the two neutrons)~\cite{Chen}
and this value (within the limits) is accepted for majority of modern
realistic $NN$ potentials~\cite{Mach_Slaus}.

Alternatively, the $a_{nn}$ values extracted from the first-type
experiments can also be divided into two categories: the first, extracted
from the Migdal-Watson approximation (MWA) for the final-state interaction 
(FSI) of two neutrons , and the
second, extracted from the exact solution of Faddeev equations for $3N$
breakup reaction. The values of $a_{nn}$ extracted from the old experiments done
until 1973 with usage of MWA have been well summarized in the review
\cite{Kuhn} and the averaged $a_{nn}$ value in \cite{Kuhn} was $-16.61\pm
0.54$ fm. It can be compared with another averaged value of $a_{nn}=-15.4\pm 0.3$~fm
found from reanalysis of data of kinematically incomplete
experiments~\cite{Tornow}. 

On the other hand, the values of $a_{nn}$ extracted with usage of the
Faddeev treatment for whole process 
are varying for different experiments from 
$a_{nn}=-16.2\pm 0.3$~fm \cite{Huhn} or $a_{nn}=-16.5\pm 0.9$~fm \cite{Witsch}
until $a_{nn}=-18.8\pm 0.5$~fm\cite{Gonz06}. It is very likely that the
rather large difference between the above values of $a_{nn}$ extracted
from {\em the same type of experiments} but using different initial energies
and different kinematical conditions is due to different contribution
of three-body forces. This contribution is not established very reliably though
the authors of the experimental results claimed that they have chosen
the three-body kinematics in such a way to minimize three-body force effects. 

However, the true origin of
three-body force is still obscure and  the above
requirement can depend upon the
three-body force operator structure which is in general still unknown.
We can add to this that the result of the Faddeev equation solution is still sensitive to
the three-body force contribution (because just this contribution explains the proper
binding energies for $^3$H and $^3$He nuclei) while the results of FSI using MWA should
be much less sensitive to the $3N$-force contribution. Thus, this difference can be a
reason for difference in sensitivity to $3N$ force contribution between the MWA and
Faddeev results for $a_{nn}$.

Thus, to extract the proper value of
$a_{nn}$ from the breakup experiments one needs to elaborate a
specific treatment which makes it possible to remove (or to minimize)
the three-body force contribution {\em objectively}, i.e. independently upon
particular structure of three-body force operator. We suggest such a
method in the present paper.\\

\section{New measurements for $a_{nn}$  in $dd$ breakup
reactions}
To get a new improved estimation for $a_{nn}$ from breakup reaction with $nn$
pair near threshold in the final state, some of the present authors made
recently a novel measurement~\cite{Konob16} using $d+d\to {}^2n_0+{}^2p_0\to
n+n+p+p$ reaction,
where $^2n_0$ and $^2p_0$ mean the near pairs of two neutrons and two
protons respectively, from whose momentum distribution the values of $a_{nn}$ can be extracted. 

The $dd$ breakup experiment was performed using a 15 MeV deuteron beam in the Skobeltsyn
Institute for Nuclear Physics in Moscow State University~\cite{Konob15}. In the measurement the
CD$_2$-target with thickness of 2 mg/cm$^2$ was used. Two protons were detected by a $\Delta E-E$ telescope at
the angle of 27$^\circ$ while a single neutron was detected at 36$^\circ$ with time-of-flight technique using the
distance of 0.79 m.
 In more detail the experimental setup scheme has been described in~\cite{Konob15}.

The resulting time-of-flight neutron spectrum was compared with the kinematic simulation results for various
values of the energy $\varepsilon_{nn}$ of the $nn$ virtual state. 
Figure 1 shows the results of such a comparison for three values of $\varepsilon_{nn}$. 
\begin{figure}[h!]
\centerline{\epsfig{file=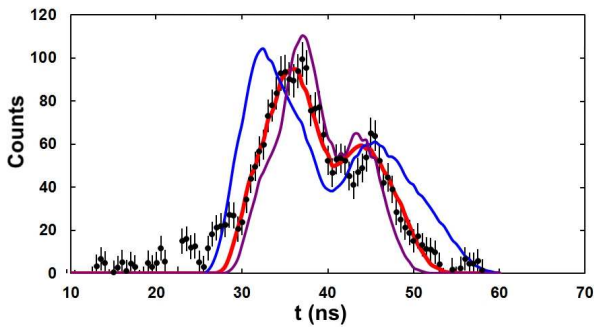,width=0.9\columnwidth}}
\caption{Experimental and simulated neutron TOF spectra for various $\varepsilon_{nn}$: 40 keV (violet), 
76 keV (red), and 160 keV (blue).}
\label{spectrum}
\end{figure}

The $\varepsilon_{nn}$-dependence of $\chi^2$ was approximated by a quadratic polynomial,
the minimum of which is achieved at $\varepsilon_{nn} = 76$~keV, $\Delta\varepsilon= \pm 6$~keV.
Then the energy of virtual level $\varepsilon_{nn}$ is related to the scattering length $a_{nn}$ and the effective range
$r_{nn}$ by the well-known equation
\begin{equation}
\frac{1}{a_{nn}}=-\left (\frac{m_n\varepsilon_{nn}}{\hbar^2}\right)^{1/2}-\frac{1}{2}r_{nn}\frac{m_n\varepsilon_{nn}}{\hbar^2},
\end{equation}
from which one gets $a_{nn}=-22.2\pm 0.6$~fm. This value very likely does include somehow the three-body
force contribution.\\ 

\section{Unified analysis for $a_{nn}$ values}
Now we use this new  $a_{nn}$ value together with previous results for $a_{nn}$ extracted from
three-nucleon breakup reaction $n+d\to nn+p$ at different energies in kinematically complete
experiments (except the result~\cite{Konob10}) for our analysis. All the $a_{nn}$ values 
corresponding to different energies (presented
in Table~1) have been taken from the dedicated experiments with full three-body
kinematics done after 1999 with usage of
fully realistic Faddeev equations, except the results \cite{Konob10,Konob16}. 
\begin{table}[h]
\caption{ The $a_{nn}$ values extracted from the kinematically complete experiments after 1999 year}
\begin{tabular}{llll}
\hline \hline
 $E_n$, MeV & $R$, fm& $a_{nn}$, fm  & Ref.\\ \hline
 15 (dd)  & 2.38  &  $-22.2\pm 0.6$ &\cite{Konob16}\\
 13  & 4.25  &  $-18.7\pm 0.6$ &\cite{Gonz99}\\  
 13  & 4.25  &  $-18.8\pm 0.5$ &\cite{Gonz06}\\
 16.6& 5.93  &  $-16.2\pm 0.3$ &\cite{Huhn}\\
 17.4& 5.16  &  $-16.5\pm 0.9$ &\cite{Witsch}\\
 19  & 5.44  &  $-17.6\pm 0.2$ &\cite{Crowe}\\
 25.3& 6.44  &  $-16.1\pm 0.4$ &\cite{Huhn}\\
 40  & 8.35  &  $-16.6\pm 1.0^a$ &\cite{Konob10}\\
\hline
\end{tabular}

$^a$ -- Here we used the corrected value of $a_{nn}$ instead of the value in \cite{Konob10} $a_{nn}=-17.9\pm 1.0$~fm 
which was obtained in the approximation of zero-range nuclear forces ($r_{nn}=0$).
\end{table}

Then we analyzed all these $a_{nn}$ values from a unified point of view of possible impact of
three-nucleon force which must be excluded. 
The criterion for the exclusion is following:
we have chosen some fixed time interval $\tau$ which corresponds to a
characteristic time for possible three-body force (e.g. which corresponds to
the average energy-exchange value $\Delta E$ in three-nucleon force operator
using the relation: $\Delta E\tau\sim\hbar$). The exact value of $\tau$ is no
matter due to evident scaling. So, if the distance $R$ between
$nn$ pair and proton (or $^2p_0$ pair in $dd$ experiment) in intermediate $3N$ (or $4N$) state 
corresponding to the incident
energy $E_n$ and the time interval $\tau$ 
is much larger than the
characteristic range of $3N$ force $r_{3N}$, i.e. if $R\gg r_{3N}$, one can
ignore the $3N$-force contribution in interpretation of the result for $a_{nn}$
in the given experiment. 

\begin{figure}[h]
\centerline{\epsfig{file=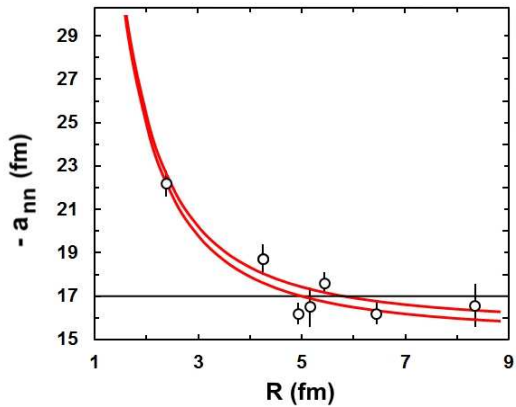,width=0.8\columnwidth}}
\caption{Dependence of the extracted $nn$ scattering length on distance $R$ between two-neutron
pair and proton (or diproton) corresponding to the initial energies 
shown in Table~I. Two curves denotes the error bars of our extrapolation
procedure.}
\label{fit}
\end{figure}
We displayed on Fig. 2  the $a_{nn}$ values extracted from experimental
data for breakup reactions $n+d\to nn+p$ (seven experiments) and $d+d\to
{}^2n_0+{}^2p_0$ (one experiment) versus distance $R$ between two-neutron
pair and proton (or $^2p_0$ pair in $dd$ experiment) corresponding to the 
initial energy for particular experiment.

It was a big surprise for us to see that all eight values of $a_{nn}$ measured in the
kinematically complete three-body experiments lie on some curve (see Fig.2) so that 
extrapolation of this curve to infinity ($R\to \infty$)
gives the asymptotic value of $a_{nn}$ which looks to be free of any three-body
force contribution:  $a^{\rm asym}_{nn}=-15.5\pm 0.5$. 
This asymptotical value for  $a_{nn}$ still includes some small magnetic interaction which is
repulsive and leads to some minor correction of $a_{nn}$. According to \cite{Mach_Slaus} this repulsive
effect can be estimated as $\Delta a_{nn}\sim -0.3$~fm. This leads to the extrapolated $a_{nn}$ 
value corrected to magnetic interaction as $a^{\rm corr}_{nn}=-15.8\pm 0.5$. 

This value is very close to
the $a_{nn}$ values, extracted for this type experiments by authors of \cite{Huhn} and our group at highest incident
energies 25 and 40 MeV respectively. It is extremely interesting also that this value of  $a_{nn}$ is rather close to
the average values summarized in \cite{Kuhn,Tornow} from  the old breakup experiments with usage of
MWA (which is not sensitive to the $3N$ force contribution).

Thus, the asymptotic value $a_{nn}=-15.8\pm 0.5$ found here gives a new impetus for
further reconsideration of the whole problem of neutron-neutron scattering
length.

\section{Charge independence symmetry in $NN$ interaction} 
Using our novel estimations for $nn$
scatterings length $a^{\rm corr}_{nn}=-15.8$~fm , it would be
interesting to reanalyze the neutron-proton scattering length, $a_{np}$, and put
the question: what are the real charge independence breaking (CIB) effects and 
can one explain the difference between $a_{np}$ and $a^{\rm corr}_{nn}$  only 
by the difference in masses of charged and neutral pions 
and the corresponding coupling constants. 

To do this, we start with well accepted value $a_{np}=-23.74$~fm and then we 
derive the $a_{nn}$ value
 using the difference between one-pion exchange potential (OPEP) in
$nn$ and $np$ pairs~\cite{Mach_Slaus}. In the first case ($nn$) it is pure
$\pi^0$-exchange while in the second case ($pn$) it is a combination  $2V_{\rm
OPE}(\pi^\pm)-V_{\rm OPE}(\pi^0)$~\cite{Mach_Slaus}. We employ the following
values for  the pion masses:  $m_{\pi^0}=134.977$~MeV, 
$m_{\pi_{\pm}}=139.570$~MeV  and respective coupling constants recommended in
\cite{Mach_Slaus}: $f_{\pi^0}^2=0.075$ $(g_{\pi^0}^2/4\pi=13.595)$ for neutral pions  and
$f_{\pi^\pm}^2=0.079$ $(g_{\pi^\pm}^2/4\pi=14.30)$ for charged pions. We use the dibaryon-model
$np$ potential (with proper OPEP) fitted to exact value $a_{np}=-23.74$~fm and
then replace $np$ OPEP with $nn$ OPEP as noted above and also replace the masses of
colliding nucleons.  This leads immediately to the value $a_{np}=-15.75$~fm
which is in a very good agreement with the value  $a^{\rm corr}_{nn}=-15.8$~fm
obtained from extrapolation of experimental values (with correction for magnetic
interaction). Thus, we see that the CIB effects in $NN$ scattering can be
explained only by differences in the potentials of one-pion exchange.

Keeping in mind the importance of this conclusion we made still another test to
remove as far as possible a model dependence of this conclusion. For this we
used the old Reid soft core (RSC) potential for singlet $np$ s-wave scattering.  We
replaced  the term in it corresponding to the  averaged OPEP with the above
correct OPEP for $np$ interaction and fitted the exact value 
$a_{np}=-23.74$~fm  using small variation of the attractive Yukawa term in RSC
potential. Then we replaced the $np$ OPEP with the proper $nn$ OPEP and have
obtained as the result $a^{RSC}_{nn}=-15.57$~fm, in good agreement with the
value for dibaryon model. Thus, our above conclusion seems to be insensitive to
the underlying $NN$-force model (but still dependent on the $\pi NN$ form factor
which regularizes the short-range behavior of OPEP~\cite{Mach_Slaus}).

The difference between the neutron-neutron scattering length $a_{nn}$ and the
``purely nuclear'' or strong-interaction  proton-proton scattering length
$a^{\rm nuc}_{pp}$ is considered as the measure of  charge-symmetry breaking.
The value $a^{\rm nuc}_{pp}=-17.3\pm 0.3$~fm \cite{Miller}  is now generally
accepted. So, the large value $a_{nn}=-18.9$~fm is greater than $a^{\rm
nuc}_{pp}$ and their difference is 1.6~fm. However, the value $a_{nn}=-15.8$~fm
found here leads to {\em the opposite sign} of charge-symmetry breaking effects
$a_{nn}-a^{\rm nuc}_{pp}=-1.5$~fm.

\section{Further implications for nuclear physics}
One of the arguments against the low value of $a_{nn}\approx -16$~fm
is the
Coulomb displacement energy between binding energies of $^3$H and $^3$He nuclei
$\Delta E_C=E_B({}^3{\rm H}) -E_B({}^3{\rm He})$. Here the well assumed point of
view is that the alternative ``large'' value $a_{nn}=-18.9\pm 0.4$~fm is
necessary to get the proper value for the Coulomb displacement energy $\Delta
E_C\simeq 740$~keV. However, this conclusion is valid only for conventional $NN$- and 
$3N$-force models. In paper~\cite{JPhys2} we have shown that within the dibaryon 
model~\cite{dibar} for $NN$ and $3N$ forces one gets the
proper value for $\Delta E_C=740$~keV using just the ``small'' value for
$a_{nn}=-16.5$~fm  {\em without any free parameters}. Thus, from this
alternative picture one can conclude that, at least, the one-to-one
correspondence between the values  $a_{nn}$ and $\Delta E_C$ is not valid and
the problem should be considered as model dependent one.

\section{Problem with description of $d(\pi^-,\gamma)nn$ reaction} 
Still another argument against the low value of $a_{nn}$ is the results for
$a_{nn}$ extracted from the stopped pion radiation capture in deuteron,
$d(\pi^-,\gamma)nn$, where two-neutron pair in $^1S_0$ final state appears at
very low energy \cite{Chen}. This reaction is considered to be free from the
$3N$-force ambiguities and due to this as a process giving the most reliable
estimation for the $a_{nn}$ value. Unfortunately, this conclusion is also not
free from the serious doubts. The detailed discussion will be presented in our
next paper, so that here we present only short outline of our arguments. In
fact, the general mechanism for the above pion reaction on deuteron can be
schematically described as following stages (see Fig.~3a): (i) the excited
two-neutron state ($C$) is emerged after the stopped $\pi^-$-absorption on
deuteron; (ii) by subsequent emission of $\gamma$-quantum this intermediate
state goes to the final $(nn)_0$ singlet state.
\begin{figure}[h]
\centerline{\epsfig{file=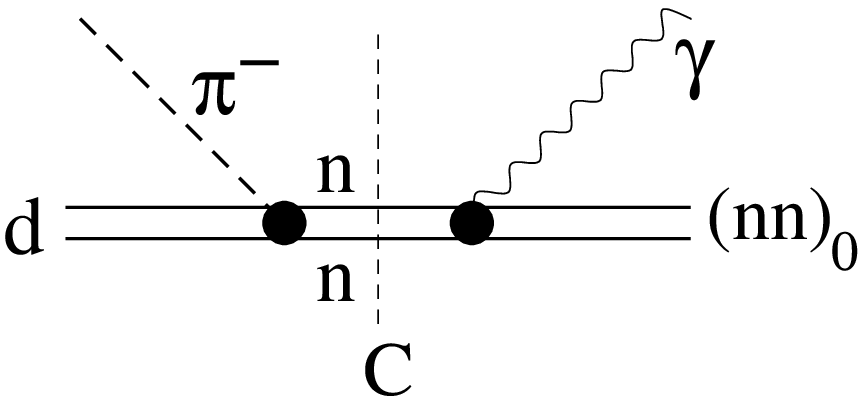,width=0.45\columnwidth} \epsfig{file=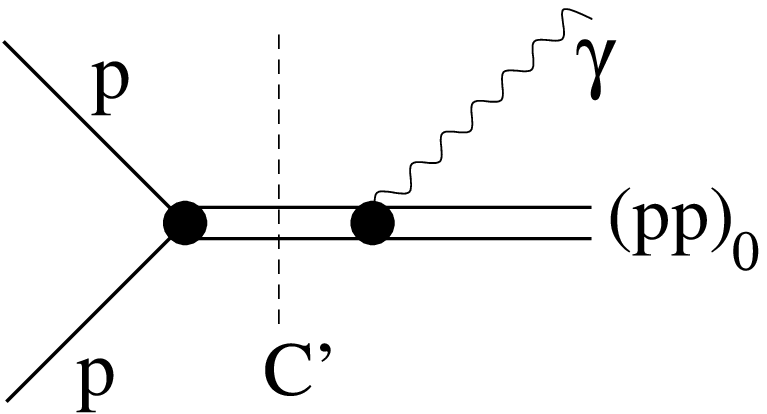,width=0.45\columnwidth}}
\caption{Two-stage mechanism for the pion radiation capture in deuteron (left) 
and $pp$ bremsstrahlung process (right).}
\label{picap}
\end{figure}

There is rather similar process $pp\to (pp)_0\gamma$ where the singlet $(pp)_0$
pair at very low energy is emerged in the final state(see Fig~3b). It is easy to see that
the final stages of both processes, starting from the intermediate excited
two-neutron $P$-wave state ($C$) and the excited
two-proton $P$-wave state ($C'$) look to be very similar. However, the best for
today theoretical description for the reaction displayed in Fig.~4b \cite{bremsstr} leads to
very serious disagreement (about 40\% for  cross section) with the accurate
experimental data, the strongest disagreement appears just in the case when the
two final protons occur in near-threshold $^1S_0$ singlet resonance state, i.e. fully similar to the 
pion capture process displayed in Fig.~4a. Moreover, the disagreement degree is increasing
with approaching the final two proton energy to the $^1S_0$ resonance region.

The true reasons for such a disagreement are still unknown but one suggests 
the very probable reason is manifestation of the of the $p$-wave dibaryon state 
in both $pp$ and $nn$ excited intermediate states
discovered recently \cite{Komarov}.

\section{Conclusion} 
In the paper we suggested some new method to reanalyze the $a_{nn}$ values
extracted from kinematically complete breakup experiments at different energies
with aim to remove somehow the three-body force contribution to the final
result. From our reanalysis we extracted the limiting value for 
$a_{nn}-15.8\pm 0.5$ which
corresponds to higher incident energy and very large separation distance between
$nn$ pair and the proton.  It should be emphasized that this $a_{nn}$ value 
is close to the magnitude of $a_{nn}$ extracted from the very numerous old
breakup experiments using MW approach.  This finding
leads immediately to the important conclusion that the charge symmetry breaking
effects in $NN$ interaction have the opposite sign as compared to the
conventional treatment. 

Our further analysis for difference between $a_{nn}$ and $a_{np}$ has shown
that the difference can be almost entirely explained by difference in OPE
interaction for $nn$ and $np$ systems. This means that the charge dependence of
the nuclear force can be attributed almost fully to the difference in OPE
interaction.

 \section*{Acknowledgments}  
 This work has been supported
 partially by the Deutsche Forschungsgemeinschaft, grant MU 705/10-1 
 and the Russian Foundation for Basic Research,
 grants No. 16-52-12005 and 16-02-00265.

\end{document}